\documentclass[11pt,article]{article}
\pdfoutput=1 

\usepackage{jheppub} 
\usepackage{color}
\usepackage[T1]{fontenc} 
\usepackage{graphicx}
\usepackage{epsfig}  
\usepackage{epsf}    
\usepackage{dcolumn}
\usepackage{bm}
\usepackage{dcolumn}
\usepackage{textcomp}
\usepackage{afterpage}
\usepackage{float}
\usepackage{subfig}
\usepackage[]{hyperref}
  \hypersetup{
  unicode=false,          
  pdftoolbar=true,        
  pdfmenubar=true,        
  pdffitwindow=true,     
  pdfstartview={FitH},    
  pdfnewwindow=true,      
  pdfcreator={RevTeX},
  colorlinks=true,       
  linkcolor=red,          
  citecolor=blue,        
  urlcolor=blue,           
  }
\usepackage{hypcap}

\newcommand{\dchisq}{\Delta \chi^2}
\newcommand{\dchisqold}{\Delta \chi^2_{\rm old}}
\newcommand{\dchisqnew}{\Delta \chi^2_{\rm new}}
\newcommand{\cninenp}{C_9^{\rm NP}}
\newcommand{\ctennp}{C_{10}^{\rm NP}}

\newcommand{\Rk}{$R_K\,$}
\newcommand{\Rkstar}{$R_{K^*}\,$}
\newcommand{\Pfive}{$P_5^\prime \,$}
\newcommand{\Bsphi}{$B_s^\phi$}

\preprint{TIFR/TH/19-6}

\title{Continuing search for new physics in $b \to s \mu \mu$ decays: \\ 
  two operators at a time}
\author[a]{Ashutosh Kumar Alok,}
\author[b]{Amol Dighe,}
\author[b]{Shireen Gangal,}
\author[c,d]{and Dinesh Kumar}

\affiliation[a]{Indian Institute of Technology Jodhpur, Jodhpur 342037, India}
\affiliation[b]{Tata Institute of Fundamental Research, Homi Bhabha Road, Mumbai 400005, India}
\affiliation[c]{National Centre for Nuclear Research, Warsaw, Poland}
\affiliation[d]{Department of Physics, University of Rajasthan, Jaipur 302004, India}
\emailAdd{akalok@iitj.ac.in}
\emailAdd{amol@theory.tifr.res.in}
\emailAdd{shireen.gangal@theory.tifr.res.in}
\emailAdd{Dinesh.Kumar@ncbj.gov.pl}

\abstract{
  The anomalies in the measurements of observables involving $b \to s \mu\mu$
  decays, namely \Rk, \Rkstar, \Pfive, and \Bsphi, may be addressed
  by adding lepton-universality-violating
  new physics contributions to the effective operators
  ${\cal O}_9, {\cal O}_{10}, {\cal O}^\prime_9, {\cal O}^\prime_{10}$.
  We analyze all the scenarios where the new physics contributes to
  a pair of these operators at a time.
  We perform a global fit to all relevant data in the $b \to s$ sector
  to estimate the corresponding new Wilson coefficients,
  $\cninenp, \ctennp, C_9^\prime, C_{10}^\prime$. 
  In the light of the new data on \Rk and \Rkstar presented in Moriond 2019, we find that the scenarios with new physics contributions to the ($\cninenp$, $C_9^\prime$) or
    ($\cninenp$, $C_{10}^\prime$) pair remain the most favored ones.
    On the other hand, though the competing scenario ($\cninenp$, $\ctennp$)
    remains attractive, its advantage above the SM reduces  significantly due to the tension that emerges between the \Rk and \Rkstar measurements with the new data. The movement of the \Rk measurement towards unity would also result in the re-emergence of the one-parameter scenario $\cninenp = -C_9^\prime$.}

\begin{document}
\maketitle
\flushbottom

\section{Introduction} 
\label{Sec:Intro}

The Standard Model (SM) of particle physics cannot be the ultimate theory
of fundamental interactions of nature. The necessity for new physics (NP)
beyond SM is indicated from multiple directions, such as the neutrino
masses, baryon asymmetry in the universe, dark matter, etc. 
Flavor physics is one of the most incisive probe of such NP, since new
particles with masses beyond the reach of current experiments can
contribute to low-energy processes through quantum corrections.
These NP effects may be measurable at dedicated flavor experiments like
LHCb~\cite{Alves:2008zz} and Belle-II~\cite{Abe:2010gxa}, as well as at multipurpose
experiments  like ATLAS~\cite{Aad:2008zzm} and CMS~\cite{Chatrchyan:2008aa}.
Deviations from the SM predictions, observed in the measurements of
processes sensitive to such effects, can provide indirect indications
of heavy particles or new interactions.
These NP effects may be quantified in a model-agnostic way,
using the language of effective field theory, by introducing additional
operators to the SM effective Hamiltonian governing the relevant processes.

Over the last few years, the rare decays of $B$ mesons, in particular
the decays induced by the quark level transition
$b \to s\, \ell^+\, \ell^-$ $(\ell=e,\,\mu)$ have already provided some such
tantalizing hints of NP. 

\begin{itemize}

\item The \Rk anomaly: The LHCb collaboration, in 2014, reported the
  measurement of the ratio $R_K \equiv  \Gamma(B^+ \to K^+ \,\mu^+\,\mu^-)/
  \Gamma(B^+ \to K^+\,e^+\,e^-)$ in the ``low $q^2$'' range
  ($1.0\, {\rm GeV}^2 \le q^2 \le 6.0 \, {\rm GeV}^2$), where $q^2$ is the 
  invariant mass-squared of the dilepton ~\cite{rk}.
  This measurement deviates from the SM value of $\simeq 1$
  ~\cite{Hiller:2003js, Bordone:2016gaq} by 2.6 $\sigma$, and is 
  an indication of lepton flavor universality (LFU) violation.
   This measurement was recently updated in Moriond 2019, including the Run-II data and an update of the Run-I analysis. The measurement of \Rk from the Run-II data is reported to be $R_K$(Run-II)$=0.928^{+0.089+0.020}_{-0.076-0.017}$,  while the combined measurement from both the runs is 
 $R_K$(new)$=0.846^{+0.060+0.016}_{-0.054-0.014}$~\cite{Rk2019}. Clearly the central value of \Rk is moving towards unity, however the discrepancy with SM has remained  $\approx 2.5\sigma$.
  
\item The \Rkstar anomaly:  The LFU violation in $b \to s\, \mu^+ \,\mu^-$
  sector  was further corroborated  by the measurement of the related
  quantity  $R_{K^*} \equiv \Gamma (B^0 \to K^{*0} \mu^+\mu^-)/
  \Gamma(B^0 \to K^{*0} e^+ e^-)$ in April 2017. The ratio $R_{K^*}$ was
  measured  in the low-$q^2$
  ($0.045 \, {\rm GeV}^2 \le q^2 \le 1.1 \, {\rm GeV}^2$),
  as well as in the central-$q^2$
  ($1.1   \, {\rm GeV}^2 \le q^2 \le 6.0 \, {\rm GeV}^2$)
  bin \cite{rkstar}. These measurements differ from the SM predictions
  of  $R_{K^*} \simeq 1$ \cite{Hiller:2003js, Bordone:2016gaq}
  by $\approx 2.4\sigma$ each. The Belle collaboration has presented their first measurements of \Rkstar in $B^0$ decays, and the world's first measurement of \Rkstar in $B^+$ decays, in Moriond 2019 \cite{rkstar2019}. These measurements, in multiple $q^2$ bins, have comparatively large uncertainties, and hence the anomaly in \Rkstar still stands at $\approx 2.4\sigma$ level.

\item The \Pfive\, anomaly: The values of the angular observable
  \Pfive\, ~\cite{DescotesGenon:2012zf,sm-angular} in $B \to K^* \mu^+ \mu^-$ decays,
  measured by the LHCb ~\cite{Kstarlhcb1,Kstarlhcb2} as well as
  ATLAS ~\cite{kstaratlas} collaboration in the
  $4.0  \, {\rm GeV}^2\le q^2 \le 6.0 \, {\rm GeV}^2$ bin,
  differ by $\approx 3.3\sigma$ ~\cite{Aebischer:2018iyb}
  from their SM prediction \cite{sm-angular}.
  This observable has also been measured by Belle and CMS
  experiments, albeit in different bins.
  While the Belle measurement
  ($4.3   \, {\rm GeV}^2 \le q^2 \le 8.68 \, {\rm GeV}^2$)
  differs from the SM by $2.6\sigma$ ~\cite{KstarBelle},
  the CMS measurement ($4.3  \, {\rm GeV}^2 \le q^2 \le 6.0 \, {\rm GeV}^2$) is
  consistent with the SM to within $1\sigma$ ~\cite{kstarcms}.
  
\item The $B_s^\phi$ anomaly: The measured value of the branching ratio
  of $B_s \to \phi \mu^+ \mu^-$~\cite{bsphilhc1,bsphilhc2} is smaller
  than the SM prediction ~\cite{Aebischer:2018iyb,Straub:2015ica} by $\approx 3.7\sigma$.

\end{itemize}

The SM predictions of \Rk and \Rkstar are theoretically clean
~\cite{Hiller:2003js, Bordone:2016gaq}, therefore the deviations of these measurements
from the SM are clear indications of NP.
On the other hand, the calculations of \Pfive\, and $B_s^\phi$
involve form factor uncertainties and undetermined power corrections
\cite{Khodjamirian:2010vf,Bobeth:2017vxj,Blake:2017fyh,Arbey:2018ics}, 
so by themselves these two anomalies cannot
be considered as unambiguous signals of NP. However, since all these four
observables are in the same ($b \to s \ell^+ \ell^-$) sector, simultaneous
anomalies observed in them should be taken seriously and addressed
within the same framework.
While the \Rk and \Rkstar anomalies could be due to NP
in $b \to s \mu^+ \mu^-$ and/or $b \to s e^+ e^-$ decays \cite{Bhatia:2017tgo,Capdevila:2017bsm,Kumar:2019qbv,DAmbrosio:2019tph},
the discrepancies in $P'_5$ and  $B_s^\phi$ can be attributed to the
presence of new physics only in  $b \to s \, \mu^+ \, \mu^-$.
Hence it would be natural to account for  all of these anomalies
by assuming new physics only in the $b \to s \mu^+ \mu^-$ sector,
which naturally breaks the LFU.
We follow this assumption throughout this work.

We analyze the above four anomalies within the framework of effective
field theory,
with the aim of gauging the effects of new operators with different Lorentz
structures that may contribute to $b \to s \mu\mu$ processes. 
While the possible Lorentz structures are vector (V), axial vector (A), 
scalar (S), pseudo-scalar (P), and tensor (T), the last three are
heavily constrained from the measurements of $B_s \to \mu \mu$ and
$b \to s \gamma$ ~\cite{Alok:2010zd,Alok:2011gv,Bardhan:2017xcc}. 
Hence in our analysis, we consider NP in the form of $V$ and $A$ operators
only. Among possible operators,
${\cal O}_9=(\bar{s} \gamma^\mu P_L b)\, (\bar{\mu} \gamma^\mu \mu)$ and
${\cal O}_{10}=(\bar{s} \gamma^\mu P_L b)\, (\bar{\mu} \gamma^\mu \gamma^5 \mu)$
already exist in the SM effective Hamiltonian, however their Wilson
coefficients (WCs) may be modified due to NP. There are also two
chirality-flipped operators,
${\cal O}^\prime_9=(\bar{s} \gamma^\mu P_R b)\, (\bar{\mu} \gamma^\mu \mu)$ and
${\cal O}^\prime_{10}=(\bar{s} \gamma^\mu P_R b)\, (\bar{\mu} \gamma^\mu \gamma^5 \mu)$,
which do not exist in the SM but may be provided by NP.
We represent the WCs of these operators by $C_9, C_{10}, C_9^\prime$ and $C_{10}^\prime$,
respectively. The NP contribution to $C_9$ and $C_{10}$ are denoted
by $\cninenp$ and $\ctennp$, respectively, i.e. $C_9 = C_9^{\rm SM} + \cninenp$ and $C_{10} = C_{10}^{\rm SM} + \ctennp$.

After the advent of the $R_{K^*}$ result in 2017, several analyses were
performed with an aim of identifying the Lorentz structure of possible NP
\cite{Capdevila:2017bsm, Altmannshofer:2017yso,DAmico:2017mtc,Hiller:2017bzc,Geng:2017svp,Ciuchini:2017mik,Celis:2017doq,Alok:2017sui,Alok:2017jgr}.
Most of these analyses showed that these anomalies,
except the low-$q^2$ bin \Rkstar measurement, may be explained by using 
a combination of $\cninenp, \ctennp, C_9^\prime$, and $C_{10}^\prime$. 
The explanation of the \Rkstar (low-$q^2$) anomaly would need the
introduction of a tensor operator~\cite{Bardhan:2017xcc},  or light $Z'$ mediators \cite{Datta:2017ezo,Altmannshofer:2017bsz}.
On the other hand, these explanations cannot help in
  resolving the other anomalies considered in this paper.
The resolution of the \Rkstar (low-$q^2$) anomaly is therefore taken to be decoupled
from that of the others, and we do not dwell on that in this paper.

The most parsimoneous solutions to the anomalies would be  the ``1D'' scenarios, where only one new WC contributes, or the values of two new WCs are related, so that there is only
one
extra parameter. The scenarios with only-$\cninenp$,
$\cninenp = - \ctennp$, or $\cninenp = - C_9^\prime$ fit the data
much better than the SM~\cite{Capdevila:2017bsm}, though the last one seems to be
disfavored since it predicts $R_K \approx 1$ \cite{Alok:2017sui}.
The above 1D scenarios can indeed be generated in several proposed new physics models that contribute to $b \to s\, \mu^+ \mu^-$ at the tree level. For example, $Z^\prime$ models with gauge couplings to leptons can generate the only-$\cninenp$ scenario \cite{Crivellin:2015lwa,Allanach:2015gkd,Altmannshofer:2016jzy}. Some leptoquark models \cite{Gripaios:2014tna,Fajfer:2015ycq,Varzielas:2015iva,Alonso:2015sja,Calibbi:2015kma,Barbieri:2015yvd,Blanke:2018sro}, and $Z^\prime$ models with loop-induced couplings or with heavy vector-like fermions \cite{{Belanger:2015nma,Boucenna:2016wpr,Boucenna:2016qad}}, can give rise to $\cninenp = -\ctennp$ scenarios. In $Z^\prime$ models with vector-like fermions and $L_\mu- L_\tau$ symmetry, the $\cninenp = -C_9{^\prime}$ scenario may be generated \cite{Altmannshofer:2014cfa}.

The ``2D'' scenarios, where NP contributes to two of the WCs, would be expected to
give much better fits to the data than the SM or the 1D fits.
The scenarios contributing to the pairs
$(\cninenp,\ctennp), (\cninenp, C_9^\prime)$ and $(\cninenp, C_{10}^\prime)$
have been shown to be able to account for all the above anomalies,
except the low-$q^2$ bin \Rkstar measurement, to a reasonable extent \cite{Capdevila:2017bsm}. Out of these scenarios, the $(\cninenp, C_9^\prime)$ may be generated in $Z^\prime$ models with couplings 
to leptons through the $L_{\mu} - L_{\tau}$ portal \cite{Altmannshofer:2014cfa}. The relative importance of these different 2D scenarios needs to
be freshly analyzed in the light of the updated \Rk and \Rkstar results.

In this paper, we analyze all the 2D scenarios, i.e. where NP
contributes to two WCs at a time in an uncorrelated manner, 
with the inclusion of the 2019 Moriond update of the \Rk and \Rkstar data.
We perform a global fit to the anomalies as well as to the related
data on observables that involve $b \to s \mu \mu$ transitions
and would be affected by the same WCs. Since all the observables
we consider are CP-conserving, we restrict the WCs to be real.
We also consider the fate of the 1D scenarios, which naturally emerge
as subsets of the relevant 2D scenarios. We focus on pointing out any
changes in the fits to the different scenarios due to the 2019 update. We also interpret these changes in terms of analytic approximations to \Rk and \Rkstar in various scenarios. 

The plan of the paper is as follows. In Sec.~\ref{sec:methodology},
we discuss the methodology adopted in our analyses. In Sec.~\ref{sec:fit},
we provide the results of our fits and discuss various 2D scenarios and their 1D sub-scenarios.
Finally, we summarize and conclude in Sec.~\ref{sec:summary}, with a
comparison among different scenarios.

\section{Methodology}
\label{sec:methodology}

We represent the effective Hamiltonian for the decay $ b \to s \mu \mu$
in the presence of new physics V and A operators by 
\begin{eqnarray}
  \mathcal{H}_{\mathrm{eff}}(b \rightarrow s \mu \mu) =
  \mathcal{H}^{\rm SM} + \mathcal{H}^{\rm VA}  \; ,
\label{heff}
\end{eqnarray}
where the SM effective Hamiltonian is 
\begin{align} \nonumber
  \mathcal{H}^{\rm SM} &= -\frac{ 4 G_F}{\sqrt{2} \pi} V_{ts}^* V_{tb}
  \bigg[ \sum_{i=1}^{6}C_i \mathcal{O}_i
    + C_7\frac{e}{16 \pi^2}[\overline{s} \sigma_{\mu \nu}
      (m_s P_L + m_b P_R)b] F^{\mu \nu} \\
    & + C_8 {\mathcal O}_8 + C_9^{\rm SM} \frac{\alpha_{\rm em}}{4 \pi}
    (\overline{s} \gamma^{\mu} P_L b)(\overline{\mu} \gamma_{\mu} \mu)
    + C_{10}^{\rm SM} \frac{\alpha_{\rm em}}{4 \pi}
    (\overline{s} \gamma^{\mu} P_L b)(\overline{\mu} \gamma_{\mu} \gamma_{5} \mu)
    \bigg] \; .
\end{align}
Here $G_F$ is the Fermi constant and $V_{ij}$ are the Cabibbo-Kobayashi-Maskawa
(CKM) matrix elements.
The Wilson coefficients $C_i$ of the four-fermi operators ${\cal O}_i$
encode the short-distance contributions to the Hamiltonian,
where the scale-dependence is implicit, i.e. $C_i \equiv C_i(\mu)$
and ${\cal O}_i \equiv {\cal O}_i(\mu)$. 
The operators ${\cal O}_i$ ($i=1,...,6,8$) contribute to these processes
through the modifications  $C_7(\mu) \rightarrow C_7^{\mathrm{eff}}(\mu,q^2)$
and $C_9(\mu) \rightarrow C_9^{\mathrm{eff}}(\mu,q^2)$, where $q^2$ is the
invariant mass-squared of the final state muon pair.
The NP effective Hamiltonian is
\begin{align} \nonumber
  \mathcal{H}^{\rm VA} &=
  -\frac{\alpha_{\rm em} G_F}{\sqrt{2} \pi} V_{ts}^* V_{tb}
  \bigg[C_9^{\rm NP} (\overline{s} \gamma^{\mu} P_L b)
    (\overline{\mu} \gamma_{\mu} \mu)
    + C_{10 }^{\rm NP} (\overline{s} \gamma^{\mu} P_L b)
    (\overline{\mu} \gamma_{\mu} \gamma_{5} \mu)  \\
    & \qquad \qquad \qquad
    + C_9^{\prime}(\overline{s} \gamma^{\mu} P_R b)(\overline{\mu} \gamma_{\mu} \mu)
    + C_{10 }^{\prime} (\overline{s} \gamma^{\mu} P_R b)
    (\overline{\mu} \gamma_{\mu} \gamma_{5} \mu) \bigg].
\end{align}
The NP effects are thus encoded in the Wilson coefficients
$C_9^{\rm NP}, C_{10}^{\rm NP},C_9^{\prime}$ and $C_{10}^{\prime}$.

While NP can in principle contribute to all the above four WCs,
we focus on those scenarios where only two of these coefficients are nonzero. 
While this restriction is somewhat arbitrary at this stage, it is possible
that symmetries of the NP at high scales can naturally make some of these
coefficients vanish.
The scenarios we consider may provide clearer insights on the role of NP
Lorentz structures, due to the smaller number of parameters involved.
We consider all six possible pairs of these coefficients, viz.
$(\cninenp, \ctennp)$, $(\cninenp, C_{9}^\prime)$, $(\cninenp, C_{10}^\prime)$,
$(\ctennp, C_{9}^\prime)$, $(\ctennp, C_{10}^\prime)$ and $(C_9^\prime, C_{10}^\prime)$.
This analysis is also naturally applicable to the scenarios where
only one of these coefficients is nonzero, or the two are linearly related,
as considered in ~\cite{Descotes-Genon:2015uva,Hurth:2016fbr,Altmannshofer:2017fio}.

For each of these pairs of WCs, we perform a global fit to the observables
that would be influenced by these WCs. Apart from the four observables
that have indicated anomalies, viz. \Rk, \Rkstar, \Pfive, \Bsphi, we also
include the constraints from
(i) the branching ratio of $B_s \to \mu\mu$ \cite{Aaij:2013aka,CMS:2014xfa,Aaboud:2018mst},
(ii) the differential branching ratios of $B^0 \to K^{*0} \mu^+ \mu^-$
~\cite{Aaij:2016flj,CDFupdate,Chatrchyan:2013cda,Khachatryan:2015isa},
$B^{+} \to K^{*+}\mu^{+}\mu^{-}$, $B^{0}\to K^{0}\mu^{+}\mu^{-}$,
$B^{+}\longrightarrow K^{+}\mu^{+}\mu^{-}$ \cite{Aaij:2014pli,CDFupdate},
and $B \to X_{s}\mu^{+}\mu^{-}$ \cite{Lees:2013nxa}
in several $q^2$ bins,
(iii) angular observables in $B^0 \to K^{*0} \mu^+ \mu^-$
~\cite{kstaratlas,kstarcms,Khachatryan:2015isa,Kstarlhcb2,CDFupdate} 
and $B^{0}_{s}\to \phi \mu^{+}\mu^{-}$ ~\cite{bsphilhc2}
in several $q^2$ bins. 
For global fits with the new data, we include the updated measurement of \Rk \cite{Rk2019} and the new measurements of \Rkstar by the Belle collaboration  (the bins $0.045 \,{\rm GeV}^{2}< q^2 < 1.1\, {\rm GeV}^2, \, 1.1 \,{\rm GeV}^{2}< q^2 < 6.0\, {\rm GeV}^2, $ and $15.0\, {\rm GeV}^{2}< q^2 < 19.0\, {\rm GeV}^2$), for $B^0$ as well as $B^+$ decays \cite{rkstar2019}.

Note that all these observables are CP-conserving, as a result we do
not expect to be sensitive to the complex nature of the new WCs.
We therefore take $C_9^{\rm NP}, C_{10}^{\rm NP},C_9^{\prime}$ and $C_{10}^{\prime}$
to be real for the sake of this article.
We perform a two-dimensional (2D) $\chi^2$ fit using the CERN minimization
code {\tt MINUIT} \cite{James:1975dr}. The  $\chi^2$ function is defined as
\begin{equation}
  \chi^2(C_i,C_j) = \big[\mathcal{O}_{\rm th}(C_i,C_j) -\mathcal{O}_{\rm exp}\big]^T \,
  \mathcal{C}^{-1} \, \big[\mathcal{O}_{\rm th}(C_i,C_j) -\mathcal{O}_{\rm exp} \big]\,.
\end{equation} 
Here $\mathcal{O}_{\rm th}(C_i,C_j)$ are the theoretical predictions of
the N=116 (122) observables before (after) the Moriond 2019 update used in the fit, while $\mathcal{O}_{\rm exp}$ are 
the experimental measurements.
The $N \times N$ total covariance matrix $\mathcal{C}$ is 
obtained by adding
the individual theoretical and experimental covariance matrices.
The values of $\mathcal{O}_{\rm th}(C_i,C_j)$ and the theoretical covariance
matrix are calculated using {\tt flavio} \cite{flavio}.
The correlations among $\mathcal{O}_{\rm exp}$ are included for the angular
observables in $B \to K^{(*)} \mu^+ \mu^-$ \cite{Kstarlhcb2}
and $B_s \to \phi \mu^+ \mu^-$ \cite{bsphilhc2}. For the branching ratio of $B_s \rightarrow \mu \mu$, we use the combined fit to $B_s \to \mu \mu$ and $B^0 \to \mu \mu$ measurements \cite{Aaij:2013aka,CMS:2014xfa,Aaboud:2018mst}, obtained by taking $B^0 \to \mu \mu$ to be SM-like \cite{Aebischer:2019mlg}.
For the other observables, we add the statistical and systematic errors in quadrature. Wherever the errors are asymmetric, 
we use the conservative approach of using the larger error on both sides of the central value.

We denote the value of $\chi^2$ in the SM by $\chi^2_{\rm SM}$, and the
best-fit value in the presence of NP by $\chi^2_{\rm bf}$. Clearly
the addition of two degrees of freedom provided by the two new WCs decreases
the $\chi^2$, and hence $\chi^2_{\rm SM} > \chi^2_{\rm bf}$. We define
$\Delta\chi^2 \equiv \chi^2_{\rm SM}-\chi^2_{\rm bf}$ for each pair of
WCs, which would enable us to quantify the extent to which a particular
combination of WCs is able to provide a better fit to the data.
For convenience of notation, we denote the value of $\dchisq$ before (after)
the 2019 update as $\dchisqold$ ($\dchisqnew$).

\section{Results and discussions}
\label{sec:fit}
\afterpage{
\begin{figure}[!p]
\includegraphics[width=0.38\textwidth]{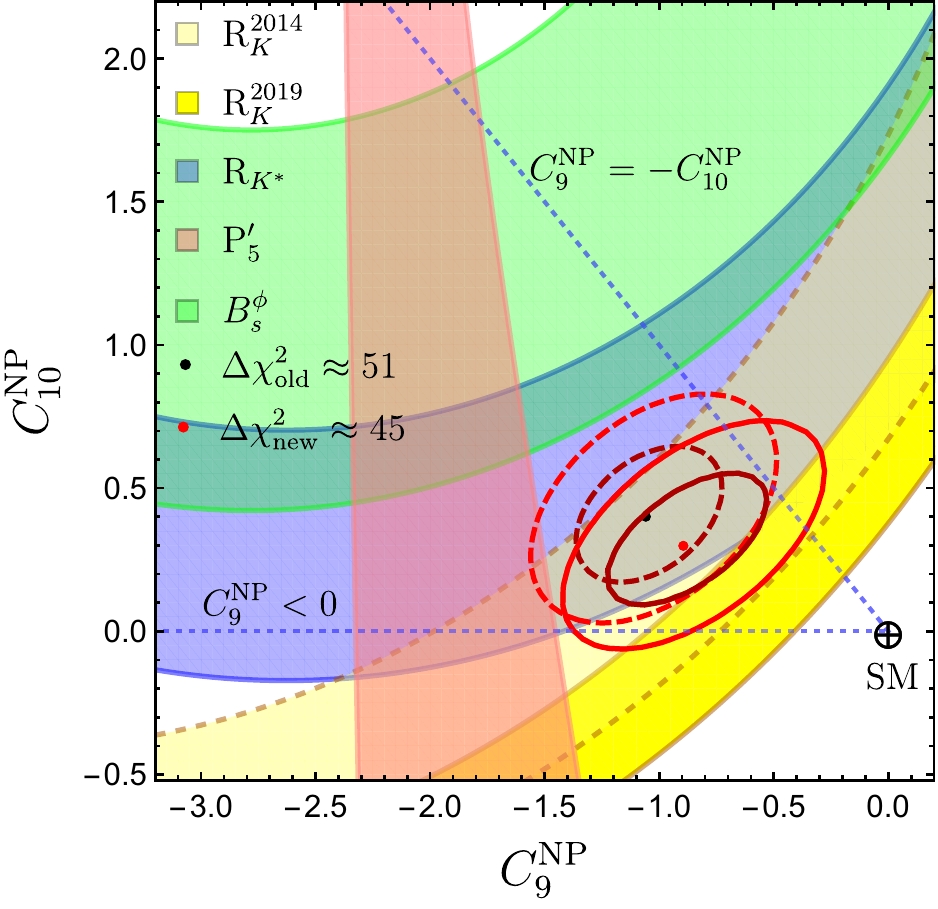}
\hspace{0.5cm}
\includegraphics[width=0.38\textwidth]{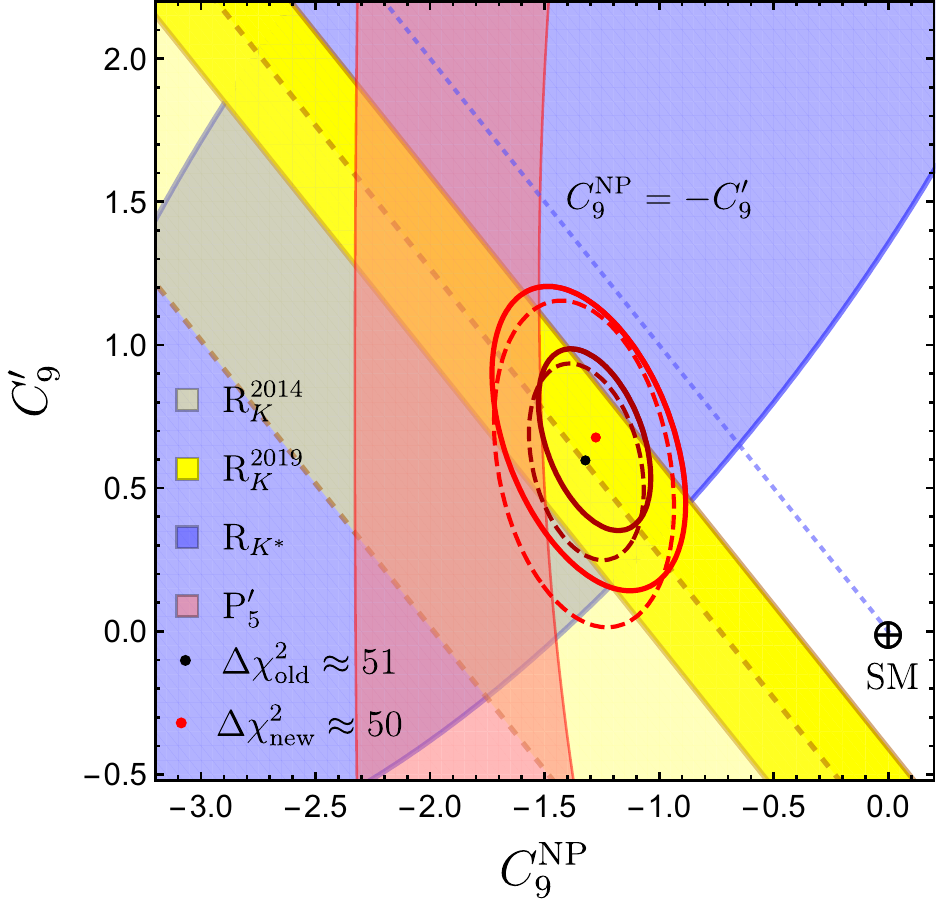} 
\\

\includegraphics[width=0.39\textwidth]{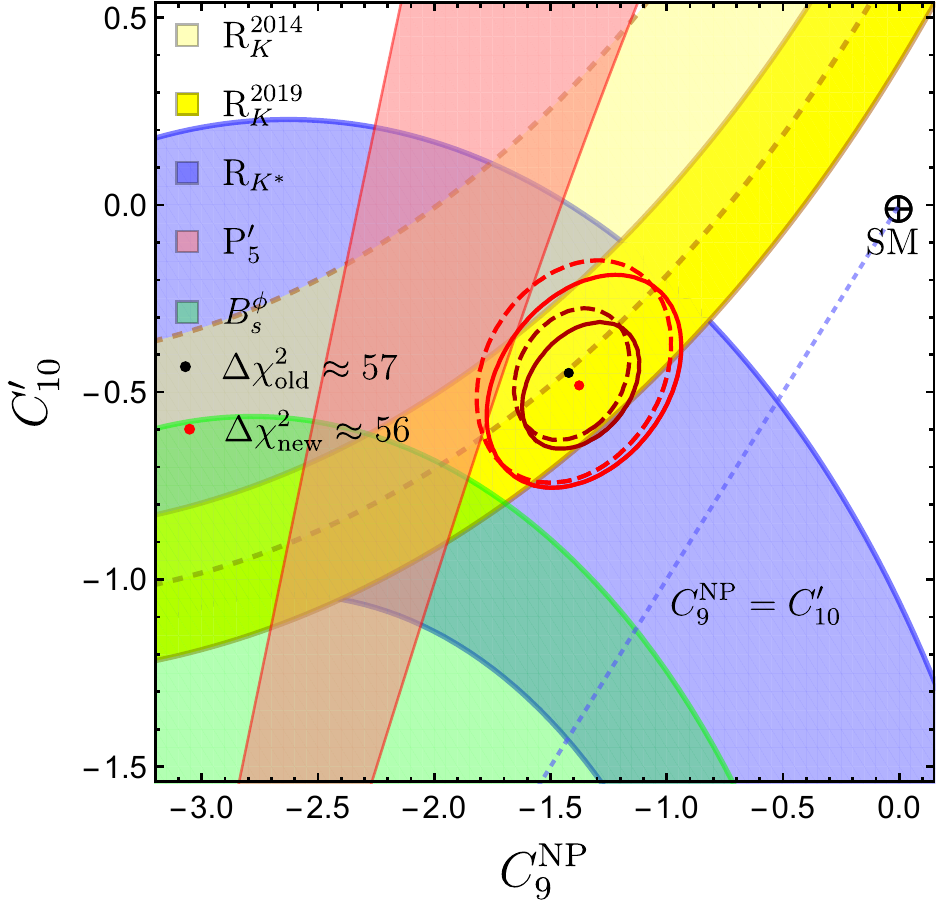}
\hspace{0.5cm}
\includegraphics[width=0.38\textwidth]{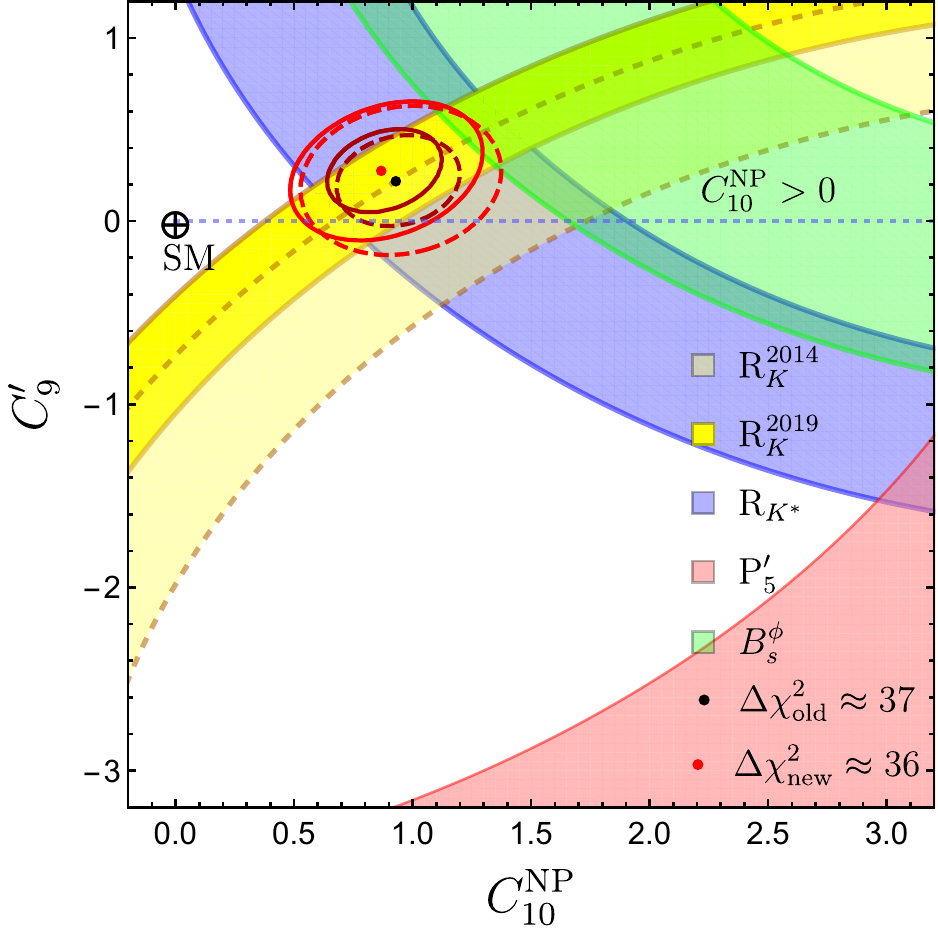} \\

\includegraphics[width=0.39\textwidth]{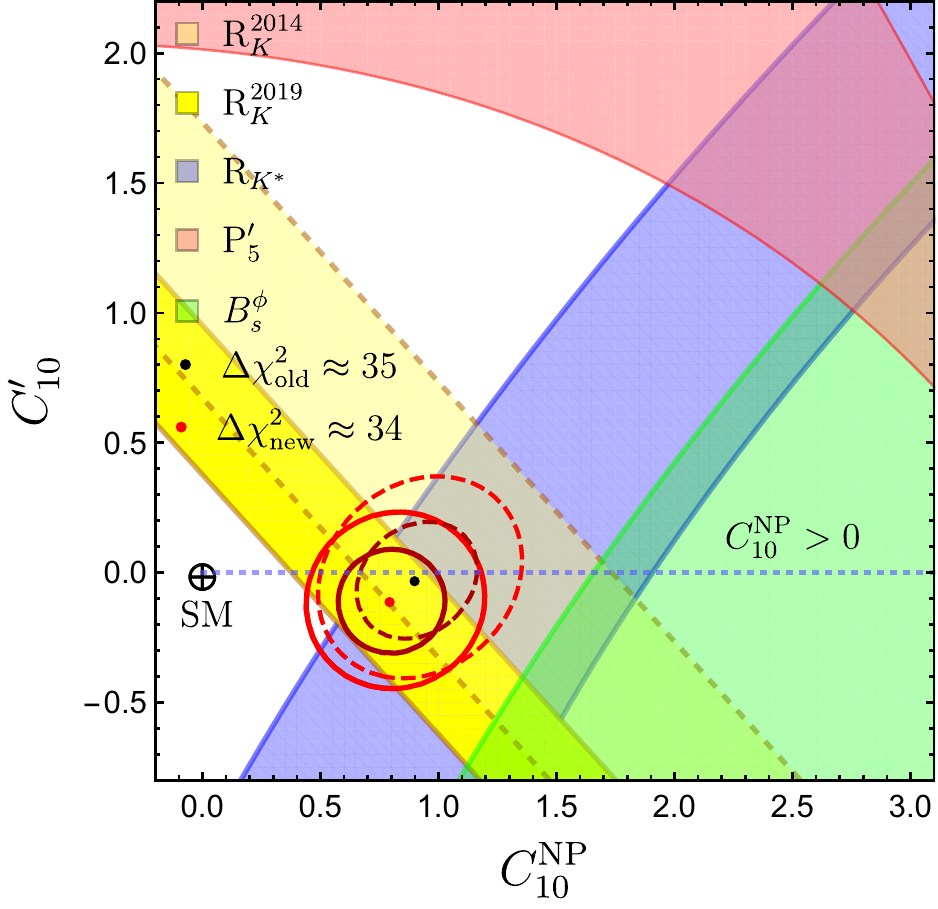}
\hspace{0.5cm}
\includegraphics[width=0.38\textwidth]{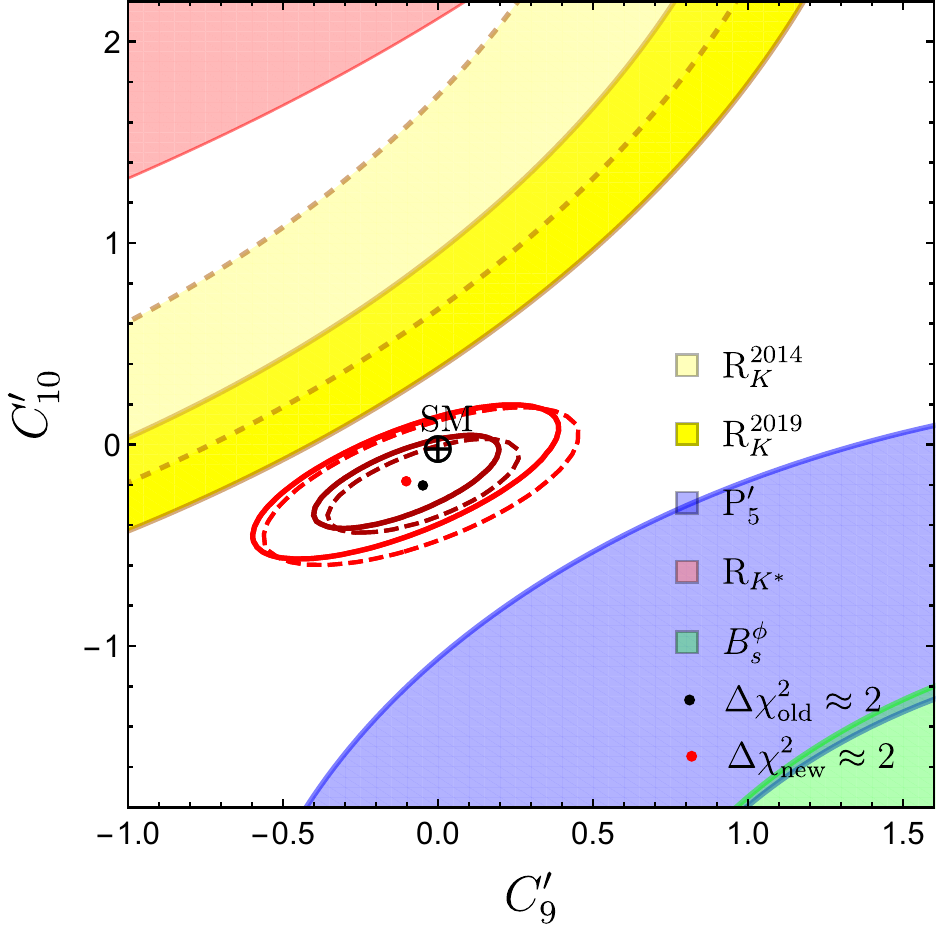}
\caption{The 1$\sigma$ allowed bands for $R_K\,(1.0\,\rm GeV^2 < q^2 < 6.0\, \rm GeV^2)$ before and after 2019 update,  $R_{K^*}$ in the central bin $\,(1.0\, \rm GeV^2 < q^2 < 6.0\, \rm GeV^2$), $P_5^\prime\,(4.0\,\rm GeV^2 < q^2 < 6.0\, \rm GeV^2)$ from ATLAS and LHCb, and \Bsphi $\equiv \mathcal{B}(B_s \to \phi \mu^+ \mu^-)$ in the range $(1.0\,\rm GeV^2 < q^2 < 6.0\, \rm GeV^2)$, for the six 2D scenarios. The 1$\sigma$ and 2$\sigma$ allowed regions from the global fit using data before (after) the 2019 \Rk update are shown by dashed (solid) contours. Specific 1D sub-scenarios that give a good fit to the data are also shown.
\label{fig:6plots}}
\end{figure}
\clearpage}

\begin{table}[t]
  \begin{center}
\begin{tabular}{|c||c|c||c|c|}
\hline\hline
Wilson Coefficient(s) & \multicolumn{2}{|c|}{Before {\tt Moriond 2019}} & \multicolumn{2}{|c|}{After {\tt Moriond 2019}} \\ \hline
  & Best fit values(s) & $\Delta\chi^2_{\rm old}$ & Best fit values(s) & $\Delta\chi^2_{\rm new}$ \\  
\hline
$C_i=0\,\,\rm (SM)$ & -- & 0 & -- & 0   \\ 
\hline \hline
1D Scenarios: & & & &\\ \hline
$C_9^{NP}$ & $-1.22 \pm 0.18$  & 42.7 & $-1.09 \pm 0.18$ & 39.0 \\ 
\hline 
$C_{10}^{NP}$ & $+0.89 \pm 0.17 $  & 34.2 & $+0.79 \pm 0.15$ & 32.3  \\ 
\hline 
$C_9^{\prime}$  & $ +0.17 \pm 0.16$  & 1.04 & $+0.09 \pm 0.15$ & 0.40 \\ 
\hline 
$C_{10}^{'}$   & $-0.22 \pm 0.12 $  & 3.06 &$-0.16 \pm 0.11$ & 1.92 \\ 
\hline 
$C_9^{NP} = C_{10}^{NP}$  & $+0.20 \pm 0.18 $   &  1.34 & $+0.20 \pm 0.17$ & 1.40  \\ 
\hline 
$C_9^{NP} = -C_{10}^{NP}$   & $-0.65 \pm 0.10 $  &  46.5 & $-0.53 \pm 0.09$ & 41.0  \\
\hline 
$C_9^{'} = C_{10}^{'}$    & $-0.20 \pm 0.16$  & 1.62 & $-0.19\pm 0.16$ & 1.51 \\ 
\hline 
$C_9^{'} = -C_{10}^{'}$    & $+0.12 \pm 0.08 $  & 2.49 & $+0.08 \pm 0.07$& 1.32 \\ 
\hline 
$C_9^{NP} = C_{9}^{'}$    & $-0.44 \pm 0.14 $  & 11.8 & $-0.35 \pm 0.12$ & 10.8 \\ 
\hline 
$C_9^{NP} = - C_{9}^{'}$   & $-1.12 \pm 0.17$  & 41.9 & $-1.12 \pm 0.17$ & 41.4  \\ 
\hline 
$C_{10}^{NP} = C_{10}^{'}$    &  $+0.37 \pm 0.13$ & 9.59 & $+0.29 \pm 0.11$ & 8.99  \\ 
\hline 
$C_{10}^{NP} = - C_{10}^{'}$   & $+0.43 \pm 0.10 $  & 22.3 & $+0.42 \pm 0.10$ & 22.3  \\ 
\hline 
$C_{9}^{NP} = C_{10}^{'}$    & $-0.68 \pm 0.12 $  & 33.5 &$-0.66 \pm 0.11$ & 32.3  \\ 
\hline 
$C_{9}^{NP} = - C_{10}^{'}$   & $-0.18 \pm 0.09 $   & 3.79 & $-0.17 \pm 0.08$ & 4.24  \\ 
\hline 
$C_{10}^{NP} = C_{9}^{'}$    & $+0.59 \pm 0.12$  & 27.3 & $+0.58 \pm 0.12$ & 27.0 \\ 
\hline 
$C_{10}^{NP} = - C_{9}^{'}$   & $+0.39 \pm 0.11 $   & 13.5 & $+0.32 \pm 0.09$ & 12.6  \\ 
\hline \hline
2D Scenarios: & & & &\\ \hline
$(C_9^{NP},C_{10}^{NP})$   & $(-1.06, +0.40)$  & 51.4 & $(-0.90,+0.30)$ & 44.7 \\  
\hline
$(C_9^{'},C_{10}^{'})$    & $(-0.05, -0.20 )$  &  2.26 & $(-0.10,-0.19)$ & 1.57 \\ 
\hline 
$(C_9^{NP},C_{9}^{'})$   & $(-1.32, +0.60)$ & 51.1 & $(-1.28,+0.68)$ & 50.3 \\  
\hline
$(C_9^{NP},C_{10}^{'})$   & $(-1.42, -0.45)$  & 57.4 & $(-1.38, -0.48)$ &  56.5\\ 
\hline 
$(C_{10}^{NP},C_{9}^{'})$   & $(+0.93, +0.22)$  & 36.7 & $(+0.87,+0.27)$ & 36.2 \\  
\hline
$(C_{10}^{NP},C_{10}^{'})$  & $(+0.90, -0.03)$ & 35.1 & $(+0.79, -0.11)$ & 33.7 \\ 
\hline\hline
\end{tabular}
\caption{Best fit values of new WCs in various 1D and 2D scenarios. The improvement over SM is quantified by $\Delta\chi^2 \equiv \chi^2_{\rm SM}-\chi^2_{\rm bf}$. For 1D scenarios, the 1$\sigma$ allowed ranges of the relevant WC are also provided. With the 2019 Moriond  update, the value of $\chi^2_{\rm SM}$ goes from 157 to 156.\label{fitreal}}
\end{center}
\end{table}

We present the results of our 2D fits in the form of contour plots
in the parameter space of the two relevant WCs, as shown in
Fig.~\ref{fig:6plots}. The six plots correspond to the six scenarios
with nonzero NP contributions to 
$(\cninenp, \ctennp)$, $(\cninenp, C_{9}^\prime)$, $(\cninenp, C_{10}^\prime)$,
$(\ctennp, C_{9}^\prime)$, $(\ctennp, C_{10}^\prime)$ and $(C_9^\prime, C_{10}^\prime)$,
respectively.
In all plots, SM corresponds to the point $(0,0)$.

In the figure, we show the 1$\sigma$ regions allowed from
the measurements of
(i) the ratio $R_{K^*}$ (central bin: $1.0$ GeV$^2 < q^2 < 6.0$ GeV$^2$),
(ii) the average of the angular observable $P_{5}^\prime$
($4.0$ GeV$^2 < q^2 < 6.0$ GeV$^2$) from the ATLAS and LHCb experiments
\cite{Aebischer:2018iyb}, and
(iii) the branching ratio $\mathcal{B}(B_s \rightarrow \phi\, \mu^+ \mu^-)$,
with bands of blue, pink, and green color, respectively.
The 1$\sigma$ allowed region of $R_K$ from the 2014 data ~\cite{rk}
and the updated 2019 data ~\cite{Rk2019} are shown by light and dark
yellow bands, respectively.
The overlaps (or lack of them) of these bands contain information
about the consistency (or tension) among different anomalies.
Note that none of these scenarios is able to account for the
measured value of \Rkstar in the low-$q^2$ bin within $2\sigma$.
So the band corresponding to this measurement is not shown in the
plots, though it contributes to the global fit. Also, the CMS results on \Pfive \cite{Khachatryan:2015isa} are not shown in the bands since they correspond to a different $q^2$-range. The new \Rkstar result from Belle \cite{rkstar2019} are also not shown, since they currently have large uncertainties. These results are, however, included in the global fit. 

Superimposed on the above bands are the $1\sigma$ and $2\sigma$ contours, shown in brown and red, respectively,  
corresponding to the global fit to all 116 (122) observables, before (after) the Moriond 2019 update. 
The contours corresponding to the data before (after) the  update
have dashed (solid) boundaries.
A comparison of these two sets of contours gives us an indication of
how the preferred parameter space in the particular NP scenario has
changed due to the 2019 update.
The superposition of these contours on the $1\sigma$ bands of
key individual measurements above allows us to check whether the
best-fit region is indeed able to account for all the anomalies.

Some of the plots also indicate the lines corresponding to 
selected scenarios with linear relations between the two WCs which give
good fits to the data.
While the viability of these 1D sub-scenarios may be judged qualitatively
from the figures, Table~\ref{fitreal}
lists the best-fit values of parameters, along with the
$\dchisqold$, $\dchisqnew$, and $1\sigma$ allowed regions for them.

Below we list some important observations that may be made for the six scenarios. Since the measurements of \Rk and \Rkstar are theoretically clean, and are expected to dominate the fits, we also try to understand the impact of new \Rk and \Rkstar measurements by using analytic approximations for \Rk and \Rkstar  (central-$q^2$) in the presence of the corresponding NP. Henceforth in this section, we shall refer to \Rkstar (central-$q^2$) simply as \Rkstar for the sake of brevity.

\subsection{The $\mathbf{(\cninenp,\,\ctennp)}$ scenario}

This scenario improves the global fit significantly as compared to the SM,
however $\dchisqnew \approx 45$ has decreased  substantially from its older value of $\dchisqold \approx 51$.
This is partly an effect of the new $R_K$ measurement having moved closer
to the SM prediction. The new measurements have also increased the
tension of the global best fit with all the four individual anomalies
marginally. This scenario still stands as one of the favored ones to
account for these anomalies.
The 1D sub-scenarios  $C_9^{\rm NP} = -C_{10}^{\rm NP}$ and  $C_{10}^{\rm NP}=0$
also continue to improve the global fit, however the extent of improvement has
 reduced for $C_9^{\rm NP} = -C_{10}^{\rm NP}$ ($C_{10}^{\rm NP}=0$) to  $\dchisq_{\rm new} \approx 41$ ($39$) with the new data, compared to
$\dchisq_{\rm old} \approx 46$ ($ 43$) from earlier.

The relatively sharp  decrease (compared to the other scenarios) in the value of $\Delta \chi^2$ after the Moriond 2019 update may be understood from the approximate functional forms \cite{Kou:2018nap}
\begin{align}
R_K = R_{K^*} &\approx 1 + 0.24\, (C_9^{\rm NP} - C_{10}^{\rm NP})\;.
\end{align}
It can be seen that the values of \Rk and \Rkstar are forced to be approximately equal in this scenario. While this was indeed the case before the update, after the update one has 
\Rk $\approx 0.85$ and \Rkstar $\approx 0.69$. Thus, a tension has emerged in the measurements of these two quantities, thereby decreasing the overall goodness of fit.

\subsection{The $\mathbf{(\cninenp,\,C_{9}^\prime)}$ scenario}

This scenario already provided a slightly better fit to the data than the
$(\cninenp, \ctennp)$ scenario, even before the 2019 update.
With the update, $\dchisqold \approx 51$ for this scenario has stayed almost the same at $\dchisqnew \approx 50$, indicating that it is still able to explain 
most of the data much better than the SM.
 Indeed, the fit is still consistent with \Rkstar and
\Pfive, while its agreement with \Rk has improved with the new data.
 The 1D sub-scenario $\cninenp=-C'_{9}$ also has continued to provide
a good fit to the data ($\dchisq_{\rm new} \approx 41$), however earlier 
it was considered to be disfavored as it predicted
\Rk $\approx 1$ \cite{Alok:2017sui}. The updated data, however, has moved
\Rk closer to unity. If this trend continues, this scenario could re-emerge as a favored NP solution.

In the $(\cninenp, C_9^\prime)$ scenario, the choices for $\cninenp$ and $C_9^\prime$ can allow \Rk and \Rkstar to vary independently: 
\begin{align}
R_K \approx 1 + 0.24\,(C_9^{\rm NP} + C_9^\prime) \,\,\,,\,\, R_{K^*} \approx 1 + 0.24\,C_9^{\rm NP} - 0.17 C_9^\prime \;.
\end{align}
No significant tension is therefore created because of the updated value of \Rk. The increase in the central value of \Rk after the update has only shifted the best fit point in the ($\cninenp, C_9^\prime$) plane to higher values of $\cninenp$ and $C_9^\prime$.  More importantly, the increase in the \Rk measurement has directly decreased the value of the combination $\cninenp + C_9^\prime$, making the 1D sub-scenario $\cninenp = -C_9^\prime$ more viable.

\subsection{The $\mathbf{(\cninenp,\,C_{10}^\prime)}$ scenario} 

This scenario was the one with the largest $\dchisqold \approx 57$ among all the
2D global fits before the update, and stays so ($\dchisqnew \approx 56$) even
with the update. It can accommodate \Rk and \Rkstar anomalies
within 1$\sigma$, and is quite close to the $1\sigma$ allowed
regions for \Pfive\, and \Bsphi.
Note that the possible 1D sub-scenarios $\cninenp=0$ or
$\cninenp = - C_{10}^\prime$ do not improve the SM fit significantly,
while  $\cninenp = C_{10}^\prime$ ($C_{10}^\prime=0$) improves it by $\dchisqnew \approx 32\, (39)$.

As far as the dependence of \Rk and \Rkstar on the NP parameters is concerned, this scenario is similar to the previous one:
\begin{align}
R_K \approx 1 + 0.24\,(C_9^{\rm NP} - C_{10}^\prime) \,\,\,,\,\, R_{K^*} \approx 1 + 0.24\,C_9^{\rm NP} + 0.17 C_{10}^\prime  \;.
\end{align}
While both these scenarios perform equally well in accounting for \Rk, \Rkstar, and \Pfive, the $(\cninenp, C_{10}^\prime)$ scenario can accommodate \Bsphi\, values closer to its measurement, and hence has a slightly better $\Delta \chi^2$ than $(\cninenp, C_9^\prime)$. The updated \Rk measurement shifts the best fit point to higher $\cninenp$ and lower $C_{10}^\prime$.

\subsection{The $\mathbf{(\ctennp,\,C_{9}^\prime)}$ scenario}

This scenario offers a moderate improvement over the SM, with
$\dchisqnew \approx 36$. 
The best fit for this scenario continues to be able to account for
the \Rk and \Rkstar anomalies to within $1\sigma$, however it cannot
explain \Pfive\, even within $2\sigma$.
The 1D sub-scenarios $\ctennp= C'_{9}$ ($\dchisqnew \approx 27$) and $C_9^\prime = 0$ ($\dchisqnew \approx 32$) offer some improvement
     over the SM,
however $\ctennp= - C'_{9}$
can only allow $\dchisqnew \approx 12$.

The approximate functional forms of \Rk and \Rkstar in this scenario are
\begin{align}
R_K \approx 1 + 0.24\,(-C_{10}^{\rm NP} + C_9^\prime) \,\,\,,\,\, R_{K^*} \approx 1- 0.24\,C_{10}^{\rm NP} - 0.17 C_9^\prime \;.
\end{align}
Since $C_9^\prime$ contributes to \Rk and \Rkstar with opposite signs, in order to have both \Rk and \Rkstar values less than unity, one would need a large value of $\ctennp$. However, such a large value of $\ctennp$ is disfavoured by $B_s \to \mu^+  \mu^-$ measurement, which is close to its SM prediction. As a result, the improvement above SM  is not significant in this scenario.

\subsection{The $\mathbf{(\ctennp,\,C_{10}^\prime)}$ scenario}

This scenario offers a moderate improvement over the SM, with
$\dchisqnew \approx 34$. 
The best fit for this scenario continues to be able to account for
the \Rk and \Rkstar anomalies to within $1\sigma$, however  it cannot
explain \Pfive\, even within $2\sigma$. 
The 1D sub-scenarios, $\ctennp= - C'_{10}$ and $C_{10}^\prime = 0$ offer  some improvement
($\dchisqnew \approx 22$ and $\dchisqnew \approx 32$, respectively) over the SM, however $\ctennp= C'_{10}$
can only allow $\dchisqnew \approx 9$.

The reason for only a moderate improvement in the goodness of fit over the SM is similar to the one in the previous scenario. Here,
\begin{align}
R_K \approx 1 + 0.24\,(-C_{10}^{\rm NP} - C_{10}^\prime) \,\,\,,\,\, R_{K^*} \approx 1 - 0.24\,C_{10}^{\rm NP} + 0.17 C_{10}^\prime \;.
\end{align}
Thus $C_{10}^\prime$ contributes to \Rk and \Rkstar with opposite signs, forcing $\ctennp$ to have unreasonably large values.

\subsection{The $\mathbf{(C_9^\prime,\,C_{10}^\prime)}$ scenario} 

This scenario is not able to offer any significant improvement over
the SM:
both $\dchisqold$ and $\dchisqnew$ are less than 3. 
As can be seen from the figure, the pairs of measurements
(\Rk, \Pfive) and (\Rkstar, \Bsphi) pull the best fit point in
almost opposite directions, thus keeping it close to the SM, without offering any solution to the anomalies.
These opposite pulls are mainly the result of \Rk and \Rkstar measurements. We have
\begin{align}
R_K \approx 1 + 0.24\,(C_9^{\prime} - C_{10}^\prime) \,\,\,,\,\, R_{K^*} \approx 1 + 0.17\,(-C_9^{\prime} + C_{10}^\prime)\;.
\end{align}
In the presence of only these two new WCs,
the values of \Rk and \Rkstar are forced in opposite directions from unity. 
As long as the measured values of \Rk and \Rkstar are
both less than unity, the allowed values of $C_9^\prime$ and $C_{10}^\prime$ will stay
small and cannot contribute to resolving both the anomalies simultaneously. The global fit will therefore stay poor.

\section{Summary and conclusions}
\label{sec:summary}

In this paper, we have explored whether pairs of new vector or
axial vector effective operators would allow us to explain the
anomalies observed in $b \to s$ decays,
namely \Rk, \Rkstar, \Pfive, and \Bsphi.
We have analyzed all the six pairwise combinations of
the NP Wilson coefficients $\cninenp, \ctennp, C_9^\prime, C_{10}^\prime$
that may contribute to the resolutions of these anomalies.
We have performed global fits to data available before and after
the Moriond 2019 update of \Rk and \Rkstar, in order to obtain the favored values of the
relevant WCs in these six scenarios. 
Our 2D global fits lead to the following observations:

\begin{itemize}

\item The two scenarios $(\cninenp, C_9^\prime)$ and $(\cninenp, C_{10}^\prime)$
continue to offer significantly better fits to the data
as compared to the SM ($\dchisqnew > 50$), even with the 2019 update to the data. Both of these best fits can account for \Rk, \Rkstar anomalies within $1\sigma$, and
\Pfive, \Bsphi\, anomalies within $2\sigma$.

\item The scenario $(\cninenp,\ctennp)$, which used to give a significantly
better fit ($\dchisqold \approx 51$) than the SM  before the 2019 update,
cannot offer as good an improvement  ($\dchisqnew \approx 45$) over the SM
after the update. Indeed it is the only 2D scenario whose $\Delta \chi^2$ has undergone such a sharp decrease after the update, compared to the other ones. The scenario is still viable, though the tensions
with individual experiments have increased with the update. The root cause of this may be traced to the approximately identical functional dependence of \Rk and \Rkstar to the two WCs, $\cninenp$ and $\ctennp$, in this scenario.

\item The scenarios $(\ctennp, C_{9}^\prime)$ and $(\ctennp, C_{10}^\prime)$ 
continue to offer only moderate improvements ($\dchisqnew \approx 35$)
over the SM. The worst scenario for explaining the anomalies turns out to be $(C'_9, C'_{10})$.  The best fit for this scenario is very close to the SM, and does not help in the simultaneous explanation of the anomalies. 

\end{itemize}
Many features of the above global fits, and the changes in these fits after the \Rk and \Rkstar update, may be understood in terms of the effect of new WCs on \Rk and \Rkstar using analytic approximations.  
Note that the anomaly in the low-$q^2$ bin of \Rkstar cannot be explained
by any of these 2D fits, as has been pointed out earlier.

These 2D fits also allow us to explore their 1D sub-scenarios where
only one new WC is nonzero, or where the two new WCs are linearly related. 
Such scenarios may be interesting not only from the point of view of smaller
number of parameters, but also because such relations may prevent
unwelcome effective operators from getting generated.
The following 1D sub-scenarios offer significant improvements above
the SM:

\begin{itemize}

\item  The $\cninenp = - C'_{9}$ scenario can give $\dchisqnew \approx 41$.
  While this was still the case before the update, it was not considered to
  be a favored scenario since it predicted $R_K \approx 1$, in conflict
  with the older data. The update has moved \Rk in the direction of unity, and
  has made this scenario more attractive.

\item The scenarios $\cninenp = -\ctennp$ ($\dchisqnew \approx 41$) and $\cninenp = C'_{10}$ ($\dchisqnew \approx 32$)
  provide moderate improvements over the SM.

\end{itemize}  
In our analysis, we have taken the data-driven approach and considered the addition of only a single, or a couple of, NP operators. While these would appear to be the most economical solutions in the language of effective field theory, they may not be always so from
the point of view of constructing a high scale theory. While reducing the high scale theory to a low scale effective theory, the desired new effective operator(s) may be necessarily accompanied by other additional effective operators with different Lorentz structures. Putting the coefficients of these effective operators to zero
is a possible way out, however the stability of such a scenario
needs to be guaranteed by a symmetry at the high scale, or  the scenario would involve some fine tuning of parameters.
Here we take the approach that having a good fit in a 2D scenario guarantees an equally good (if not better) fit in the space with more than two NP parameters. The favored scenarios that have emerged with the updated data could help in narrowing down possible NP models and guiding constructions of models beyond the current paradigm.


\bigskip
\noindent
{\bf Acknowledgements}: We would like to thank Diptimoy Ghosh, Jacky Kumar and Gagan Mohanty for useful discussions. The work of DK is partially supported by the National Science Centre (Poland) under the
research grant No. 2017/26/E/ST2/00470.

\appendix
\section{Updates after new measurements presented in Moriond 2021}
In this appendix, we update the results  in the main text obtained
using $b \to s\, \mu^+ \,\mu^-$ measurements till 
March 2019. Between March 2019 and March 2021, the LHCb collaboration has
updated the following measurements:
\begin{enumerate}
\item {\it Angular analysis of $B^0 \to K^{*0} \mu^+\mu^-$:} In March 2020, the LHCb collaboration 
updated the angular analysis of $B^0 \to K^{*0} \mu^+\mu^-$
by using a data set corresponding to an integrated luminosity of 4.7 ${\rm fb^{-1}}$ of pp collisions
 \cite{Aaij:2020nrf}. The updated measurement reinforced the tension seen between the previous LHCb
results and the SM predictions. However, the discrepancy in the angular observable \Pfive  
 in the $4.0  \, {\rm GeV}^2\le q^2 \le 6.0 \, {\rm GeV}^2$ bin reduced slightly (in comparison to the previous measurement
\cite{Kstarlhcb2}). In our updated fit, we have replaced the previous LHCb measurements of the angular observables  in $B^0 \to K^{*0} \mu^+\mu^- $ \cite{Kstarlhcb2} by the  updated values listed in Table I of Ref. \cite{Aaij:2020nrf}.

\item {\it Angular analysis of $B^+ \to K^{*+} \mu^+\mu^-$:} The  first measurement using data on the 
$CP$-averaged angular observables in $B^+ \to K^{*+} \mu^+\mu^-$ by the LHCb collaboration (9 ${\rm fb^{-1}}$ of pp collisions) was presented in December, 2020 \cite{Aaij:2020ruw}. Deviations from the SM predictions, similar to those in the angular observables of $B^0 \to K^{*0} \mu^+\mu^-$, were observed. We include the values of angular observables given in the Table II of Ref. \cite{Aaij:2020ruw} in our updated analysis.

\item {\it The \Rk anomaly:} Very recently, in Moriond 2021  (March 2021), the LHCb collaboration has reported the most precise measurement of \Rk \cite{Aaij:2021vac} till date. The measured value, $0.846^{+0.044}_{-0.041}$ in the 
  $1.0\, {\rm GeV}^2 \le q^2 \le 6.0 \, {\rm GeV}^2$ bin, deviates from the SM prediction at the level of 3.1 $\sigma$. The central value of \Rk has remained unchanged, however the errors have reduced by almost 30\%.
  
  \item {\it The branching ratio of  $B_s \to \mu \mu$:} In Moriond 2021, the LHCb collaboration  has also announced the most precise single experiment measurement of the branching ratio of $B_s \to \mu \mu$ to date \cite{bsmumu}. The measured value of $\left(3.09^{+0.46+0.15}_{-0.43-0.11} \right) \times 10^{-9}$, obtained after using full LHCb Run 1 and Run 2 data set  is consistent with the SM prediction of $(3.66\pm 0.14) \times 10^{-9}$ \cite{Beneke:2019slt}. The results in the main text were obtained using the combined fit to $B_s \to \mu \mu$ and $B^0 \to \mu \mu$  measurements of LHCb, CMS and ATLAS available till March 2019 \cite{Aaij:2013aka,CMS:2014xfa,Aaboud:2018mst,Aebischer:2019mlg}. In our updated fit, we use the combination of the updated measurements, as obtained recently \cite{Geng:2021nhg,Angelescu:2021lln}.

\end{enumerate}

\begin{figure}[htbp]
\includegraphics[width=0.32\textwidth]{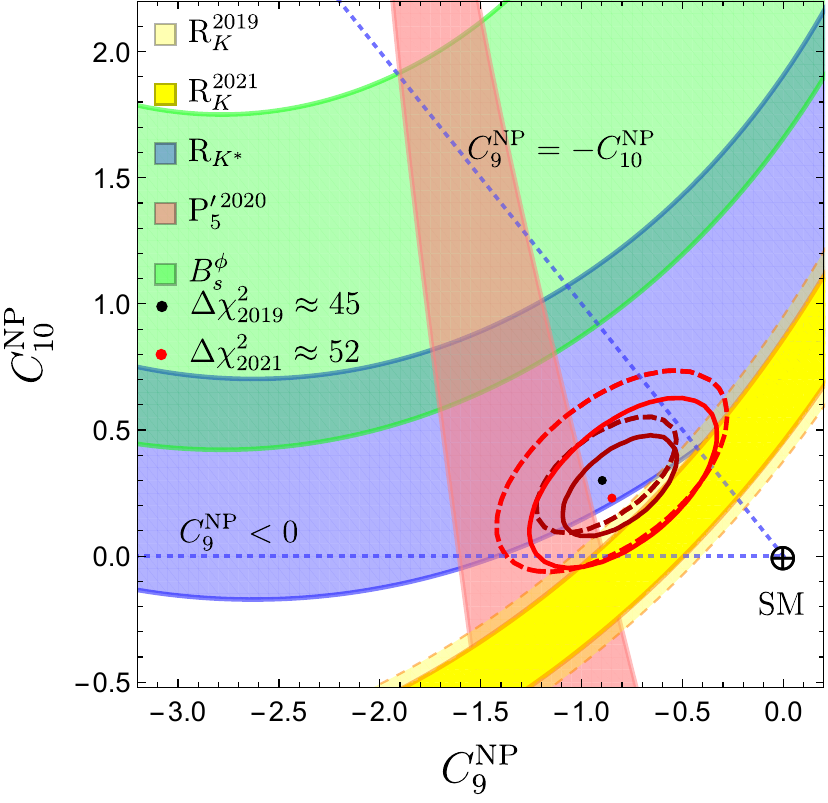}
\includegraphics[width=0.32\textwidth]{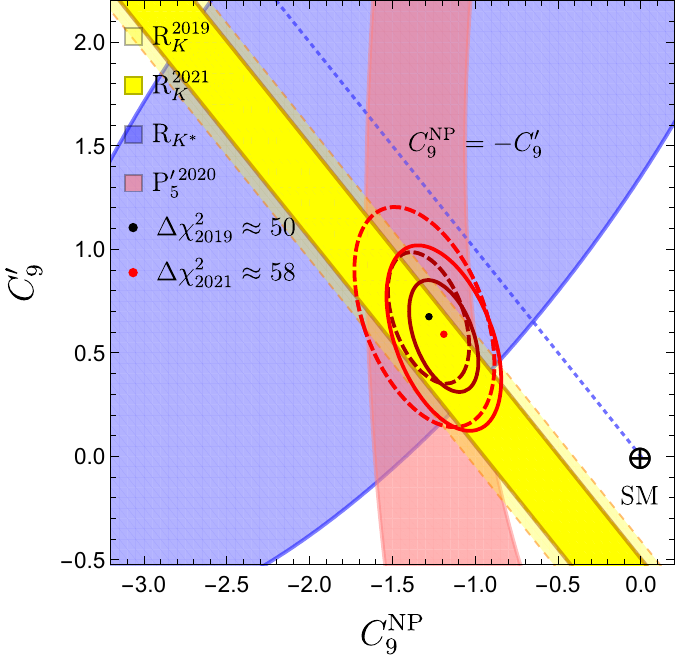}
\includegraphics[width=0.32\textwidth]{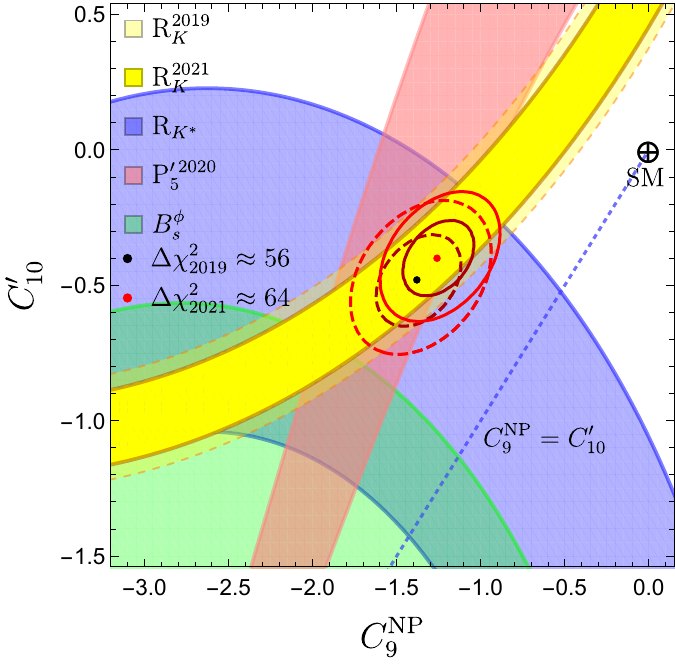}
\caption{NP scenarios $(\cninenp,\ctennp)$, $(\cninenp, C_9^\prime)$ and $(\cninenp, C_{10}^\prime)$ after Moriond 2021 updates. We denote the value of $\Delta \chi^2$ after Moriond 2019 (2021) as $\Delta \chi^2_{2019}$ ($\Delta \chi^2_{2021}$). 
\label{fig:6plots-1}}
\end{figure}

\begin{table}[t]
  \begin{center}
\begin{tabular}{|c||c|c||c|c|}
\hline\hline
Wilson Coefficient(s) & \multicolumn{2}{|c|}{After {\tt Moriond 2019}} & \multicolumn{2}{|c|}{After {\tt Moriond 2021}} \\ \hline
  & Best fit value(s) & $\Delta \chi^2_{2019}$ & Best fit value(s) & $\Delta \chi^2_{2021}$ \\  
\hline
$C_i=0\,\,\rm (SM)$ & - & 0 & &   \\ 
\hline \hline
1D Scenarios: & & & &\\\hline		
$C_9^{NP}$ & $-1.09 \pm 0.18$  & 39.0 & $-1.01 \pm 0.15$ & 48.34 \\ 
\hline 
$C_{10}^{NP}$ & $0.79 \pm 0.15 $  & 32.3 & $0.71 \pm 0.13$ & 37.87  \\ 
\hline 
$C_9^{'}$  & $ 0.09 \pm 0.15$  & 0.40 & $-0.05 \pm 0.13$& 0.17 \\ 
\hline 
$C_{10}^{'}$   & $-0.16 \pm 0.11 $  & 1.92 &$-0.06 \pm 0.10$ &0.28  \\ 
\hline 
$C_9^{NP} = C_{10}^{NP}$  & $0.20 \pm 0.17 $   &  1.40 &$0.16 \pm 0.15$ &1.23  \\ 
\hline 
$C_9^{NP} = -C_{10}^{NP}$   & $-0.53 \pm 0.09 $  &  41.0 & $-0.49 \pm 0.07$& 49.43  \\
\hline 
$C_9^{'} = C_{10}^{'}$    & $-0.19 \pm 0.16$  & 1.51 &$-0.15\pm 0.13$ & 1.28 \\ 
\hline 
$C_9^{'} = -C_{10}^{'}$    & $0.08 \pm 0.07 $  & 1.32 & $0.01 \pm 0.06$& 0.02 \\ 
\hline 
$C_9^{NP} = C_{9}^{'}$    & $-0.35 \pm 0.12 $  & 10.8 &$-0.38 \pm 0.10$ & 18.33 \\ 
\hline 
$C_9^{NP} = - C_{9}^{'}$   & $-1.12 \pm 0.17$  & 41.4 & $-1.03 \pm 0.15$& 44.55  \\ 
\hline 
$C_{10}^{NP} = C_{10}^{'}$    &  $0.29 \pm 0.11$ & 8.99 & $0.31 \pm 0.09$& 15.01  \\ 
\hline 
$C_{10}^{NP} = - C_{10}^{'}$   & $0.42 \pm 0.10 $  & 22.30 & $0.34 \pm 0.08$& 20.32  \\ 
\hline 
$C_{9}^{NP} = C_{10}^{'}$    & $-0.66 \pm 0.11 $  & 32.3 &$-0.54 \pm 0.10$ & 29.89  \\ 
\hline 
$C_{9}^{NP} = - C_{10}^{'}$   & $-0.17 \pm 0.08 $   & 4.24 & $-0.20 \pm 0.07$& 8.60  \\ 
\hline 
$C_{10}^{NP} = C_{9}^{'}$    & $0.58 \pm 0.12$  & 27.0 &$0.52 \pm 0.10$ & 27.24 \\ 
\hline 
$C_{10}^{NP} = - C_{9}^{'}$   & $0.32 \pm 0.09 $   & 12.6 & $0.32 \pm 0.08$ &19.07  \\ 
\hline \hline
2D Scenarios: & & & &\\ \hline
$(C_9^{NP},C_{10}^{NP})$   & $(-0.90, 0.30)$  & 44.70 &(-0.82, 0.27) & 52.59 \\  
\hline
$(C_9^{'},C_{10}^{'})$    & $(-0.10, -0.19) $  &  1.57 & (-0.18, -0.14)& 1.38   \\ 
\hline 
$(C_9^{NP},C_{9}^{'})$   & $(-1.28, 0.68)$ & 50.30 &(-1.19, 0.59) & 58.24  \\  
\hline
$(C_9^{NP},C_{10}^{'})$   & $(-1.38, -0.48)$  & 56.50 & (-1.26, -0.40) & 63.86  \\ 
\hline 
$(C_{10}^{NP},C_{9}^{'})$   & $(0.87, 0.27)$  & 36.20 & (0.80, 0.24) & 40.73  \\  
\hline
$(C_{10}^{NP},C_{10}^{'})$  & $(0.79, -0.11)$ & 33.70 & (0.71, -0.04) & 38.02  \\  
\hline\hline
\end{tabular}
\caption{The best fit values of new WCs in various 1D and 2D scenarios. Here $\Delta\chi^2 = \chi^2_{\rm SM}-\chi^2_{\rm bf}$ where $\chi^2_{\rm bf}$ is the $\chi^2$ at the best fit point and $\chi^2_{\rm SM}$  corresponds to the SM. Note that $\chi^2_{\rm SM}$  = 156 (199) for the 2019 (2021) data set. }
\label{fitreal-1}
 \end{center}
\end{table}

The fit results obtained after including all above mentioned updates are presented 
in  Fig.~\ref{fig:6plots-1} and Table~\ref{fitreal-1}. The number of observables has increased from  122 to 154 after including new measurements of angular observables in $B^+ \to K^{*+} \mu^+\mu^-$. Further, the results after Moriond 2019 were obtained using {\tt flavio}-1.1,  whereas for 2021 results, we use {\tt flavio}-2.2.0. Moreover, for the convenience of notation, we denote the value of $\Delta \chi^2$ after Moriond 2019 (2021) as $\Delta \chi^2_{2019}$ ($\Delta \chi^2_{2021}$).

Below we list some important observations that are evident from Fig.~\ref{fig:6plots-1} and Table~\ref{fitreal-1}:

\begin{itemize}
\item The value of $\chi^2_{\rm SM}$ has increased from 156 to 199. This is partly due to the increase in the number of observables, but also due to the new measurements  of \Rk \cite{Aaij:2021vac} whose deviation from SM is now at 3.1 $\sigma$ level.

\item The value of $\Delta \chi^2$ for the three 2D NP solutions 
 $(\cninenp,\ctennp)$, $(\cninenp, C_9^\prime)$ and $(\cninenp, C_{10}^\prime)$ has increased from their  older values $\Delta \chi^2_{2019} \approx$ (45, 50, 57) to $\Delta \chi^2_{2021} \approx$ (53, 58, 64). 
 
\item For 1D scenarios $C_9^{NP}$, $C_9^{NP}= - C_{10}^{NP}$ and $C_9^{NP} = - C_{9}^{'}$, the values of $\Delta \chi^2$ have increased from their  older values $\Delta \chi^2_{2019} \approx$ (39, 41, 41)  to $\Delta \chi^2_{2021} \approx$ (48, 49, 45). 
\end{itemize}

To summarize, these updates have resulted in an overall increase of tension with the SM. The preferred 2D and 1D NP scenarios, as obtained after Moriond 2019, have remained the same. The 2D scenario $(\cninenp, C_{10}^\prime)$ provides a better fit to the data in comparison to the other two viable scenarios. For the 1D fits, after Moriond 2019,  $\cninenp = - C'_{9}$ and $C_9^{NP}= - C_{10}^{NP}$ scenarios provided a marginally better fit in comparison to $C_9^{NP}$. However, after Moriond 2021, $C_9^{NP}$ and $C_9^{NP}= - C_{10}^{NP}$ scenarios accommodate $b \to s\, \mu^+ \,\mu^-$ data slightly better than  the  $\cninenp = - C'_{9}$ solution.

\end{document}